\documentclass[...]{PoS}
\usepackage{caption}
\usepackage{graphicx}
\usepackage{subcaption}
\usepackage{subfig}

\title{A Novel Method for Detecting Extended Sources with VERITAS}

\ShortTitle{A Novel Method for Detecting Extended Sources with VERITAS}

\author{\speaker{Joshua V. Cardenzana} for the VERITAS Collaboration\thanks{veritas.sao.arizona.edu}\\
        Department of Physics \& Astronomy, Iowa State University, Ames, IA 50011, USA\\
        E-mail: \email{jvcard@iastate.edu}}

\abstract{
The most commonly used techniques for estimating the background contribution in
IACT data analysis are the ring background model and the reflected
region methods. However, these two techniques are poorly suited for
analyses of sources with extensions comparable to the detector's field
of view (greater than $\sim$1$^{\circ}$). Nearby pulsar wind nebulae,
supernova remnants interacting with molecular clouds, and dark matter
signatures from galaxy clusters are just a few potentially highly
extended source classes. A three dimensional maximum likelihood
analysis is in development that seeks to resolve this issue for data
from the VERITAS telescopes. The technique incorporates relevant
instrument response functions to model the distribution of detected
gamma-ray like events in two spatial dimensions. Additionally, we
incorporate a third dimension based on a gamma-hadron discriminating
parameter. The inclusion of this third dimension significantly
improves the sensitivity of the technique to highly extended
sources. We present this promising technique as well as
systematic studies demonstrating its potential for revealing sources
of large extent in VERITAS data.
}

\FullConference{The 34th International Cosmic Ray Conference \\
                 30 July- 6 August, 2015\\
                 The Hague, The Netherlands}

\begin{document}
\section{Introduction}
Imaging atmospheric Cherenkov telescopes (IACTs) have had great success in
detecting sources at TeV energies. However, the sources typically
detected are point sources or exhibit only moderate extensions (radial
extension $\lesssim$ 0.5$^{\circ}$). Other instruments with larger
fields of view have detected sources with very large extensions ($\gtrsim$1$^{\circ}$)
including the cocoon detected by \textit{Fermi}-LAT in the Cygnus
region, believed to be the result of freshly accelerated
cosmic-rays confined within a cavity in the Cygnus X star-forming
region \cite{Ackermann25112011}. The morphology of the cocoon was
characterized using a Gaussian with $\sigma$ = 2.0$^{\circ}$ $\pm$
0.2 making it highly extended. Milagro also reported evidence of
extended emission (intrinsic full width at half maximum of 2.6$^{\circ}$$^{+0.7}_{-0.9}$)
\cite{2009ApJ...700L.127A} coincident with the \textit{Fermi}-LAT
Geminga pulsar \cite{0004-637X-720-1-272}. This extended source,
identified as MGRO J0632+17, is thus potentially a pulsar wind nebula
(PWN). Additionally, dark matter signatures from  galaxy clusters can
extend over regions larger than a degree
\cite{1475-7516-2013-02-028}. Detecting these sources requires an
analysis which can account for all of the emission in the region
simultaneously. This is not a trivial task given the analysis
techniques used in IACT data analysis. Standard methods for
estimating background emission for these instruments are either the
ring background model (RBM) or the reflected-region model
\cite{2007A&A...466.1219B}.

The RBM analysis estimates the level of background for a particular
region of interest (\textit{ON} region) by using an annulus, or ring, around the
region. In order to obtain an accurate measure of the background the ratio
of the \textit{ON} region to the ring region is typically on the
order of $\sim$1/7. The reflected-region model requires observations to be taken
offset from the source position. Suitable \textit{OFF} regions are
chosen at positions in the field of view which have similar offsets from the tracking
position as the established \textit{ON} region. For
sources with extensions >0.5$^{\circ}$, no \textit{OFF} regions can be
determined and a larger observing offset is needed, reducing
sensitivity to the source. Ultimately, the field of view of the instrument limits the size of source which can be observed with this technique. In the RBM highly
extended sources may require a larger \textit{ON} than \textit{OFF}
region, resulting in a poorly determined background. These examples for the RBM and
reflected-region analyses of sources with large extensions are
predicated on the assumption that the morphology of the source is
known. For the most extended sources the morphology is rarely
understood with absolute certainty. In this case, determining suitable
\textit{OFF} regions becomes very difficult and a certain amount of
self subtraction due to source contamination in the \textit{OFF}
region is almost guaranteed. Thus, a different analysis technique is
necessary to detect these sources in VERITAS data.
\section{VERITAS}
\label{AboutVERITAS}
The Very Energetic Radiation Imaging Telescope Array System (VERITAS)
is an array of four 12-meter IACTs located at the Fred Lawrence Whipple
Observatory (FLWO) in southern Arizona (31$^{\circ}$40' N,
110$^{\circ}$57' W,  1.3km a.s.l.). The instrument covers an energy
range from 85 GeV to > 30 TeV with an energy resolution between 15 -
25\%. VERITAS is capable of detecting a point source with 1\% the Crab
nebula flux in $\sim$25 hours and has an angular resolution (68\%
containment radius) of <0.1$^{\circ}$ at 1 TeV.

\section{VERITAS Maximum Likelihood Method}
\label{VERITASMaximumLikelihoodMethod}
In order to tackle extended sources in VERITAS data, a three dimensional maximum
likelihood method (MLM) is in development. In general, the method of maximum
likelihood works by modeling the expected distribution of events in
data. A likelihood value \textit{L} can then be computed from these models
according to
\begin{equation}
\label{eq:BaseLikelihood}
L(\mathbf{ \vec{s} } ) = \prod \limits_{i=1}^{d} p ( \mathbf{\theta} _{\textit{i}} | \mathbf{ \vec{s} } )
\end{equation}
where \textit{p} is the model, \textit{i} indexes individual data
events, $\theta$ is a set of observables which represent the
data, and $\mathbf{\vec{s}}$ is a set of free parameters chosen to maximize
\textit{L}. Typically, the log of the likelihood is computed as this turns
the product into a sum and the computation is simplified. 

The VERITAS MLM models consist of both a two dimensional spatial
model and a mean scaled width model (MSW) for both the $\gamma$-ray (source)
and background components. In this way the full
log-likelihood for a single observation becomes
\begin{equation}
\label{eq:VERITASLikelihood}
log(L(\mathbf{ \vec{s} } )) = N_{obs}log(N_{exp})-N_{exp}+\sum \limits_{i=1}^{d}
log\left[S_{src}(\mathbf{\vec{r}}_{\textit{i}}|\mathbf{ \vec{s} })\cdot
  W_{src}(w_{\textit{i}}|\mathbf{ \vec{s} })+S_{bkgd}(\mathbf{\vec{r}}_{\textit{i}})\cdot W_{bkgd}(w_{\textit{i}})\right]
\end{equation}
To correlate the spatial source
model (\textit{S$_{src}$}) with the MSW source model
(\textit{W$_{src}$}), the product of the two distributions is taken
(similarly for the background component). An additional term is
included in the likelihood to model the total expected number of
events from both signal and background. The expected number of events
in a dataset is taken to be Poisson distributed with expected events
\textit{N$_{exp}$} and observed events \textit{N$_{obs}$}. The first
two terms in Eq. \ref{eq:VERITASLikelihood} are based on this
assumption. The data is treated as unbinned in the spatial and MSW
dimensions and coarsely binned in reconstructed energy. The free
parameters are spectral parameters associated with the source being modeled. 

It follows that to analyze data from multiple
observing positions, each field can be independently modeled and the
resulting log-likelihoods summed to get a likelihood
for the entire observation set simultaneously. Since all models share
the same free parameters
(which are independent of the observing conditions between fields)
this can be extended to observations which cover a range of observing
positions, detector configurations, or even to connect observations
from different detectors.

\section{Defining the Input Models}\label{ModelCreation}
In this section, the various models which represent the expected
distribution of the data, and how they are
derived is detailed. Specific attention will be paid to the source spatial model
in Section \ref{SourceSpatialModel} outlining the details of how we
incorporate the various instrument response functions (IRFs) from
VERITAS. Each model component is derived from either data (primarily
for modeling background) or simulated $\gamma$-ray air-showers
(for source models). In both cases, events were processed using
standard medium cuts for all variables with the exception of MSW in
which a larger range was used to better characterize the background models.

\subsection{Background Spatial Models}
The background spatial models represent the expected distribution of
$\gamma$-ray-like events across the field of view. They are derived 
from observations taken on weak blazars and dwarf
spheroidal galaxies. Data taken in the galactic plane is not used to prevent
possible contamination from diffuse $\gamma$-rays. These data are first
transformed into the spatial coordinate system used in the
analysis and then binned into 0.01$^{\circ}$ spatial bins. A
radial acceptance distribution is generated in \textit{r}$^{2}$ by
summing all bins within a range of offsets
from the field of view center. This radial acceptance is extrapolated
into two dimensions to form the spatial model associated with the
expected background.

Because data is used to derive the background spatial models there are two
main effects which must be accounted for. The first effect is contamination
from any potential source within the field of view which results
in an overestimate of the background at the offset of the
source. Secondly, stars in the field of view can cause deficits in
the surrounding region. To alleviate these effects all bins within
a set angular distance\footnote{The size of the exclusion region for 
  models in this paper is 0.4$^{\circ}$, however this value is still
  undergoing optimization.} of known (or potential) sources and bright
stars\footnote{\textit{bright stars} are defined as those with B magnitude
  8 or below.} are excluded. To correct for the excluded regions, each bin of the
radial acceptance curve is weighted by the relative number of bins which actually contribute to it.
\subsection{Source Spatial Model}\label{SourceSpatialModel}
\begin{figure}
   \centering
   \begin{subfigure}[b]{0.4\textwidth}
      \includegraphics[width=\textwidth]{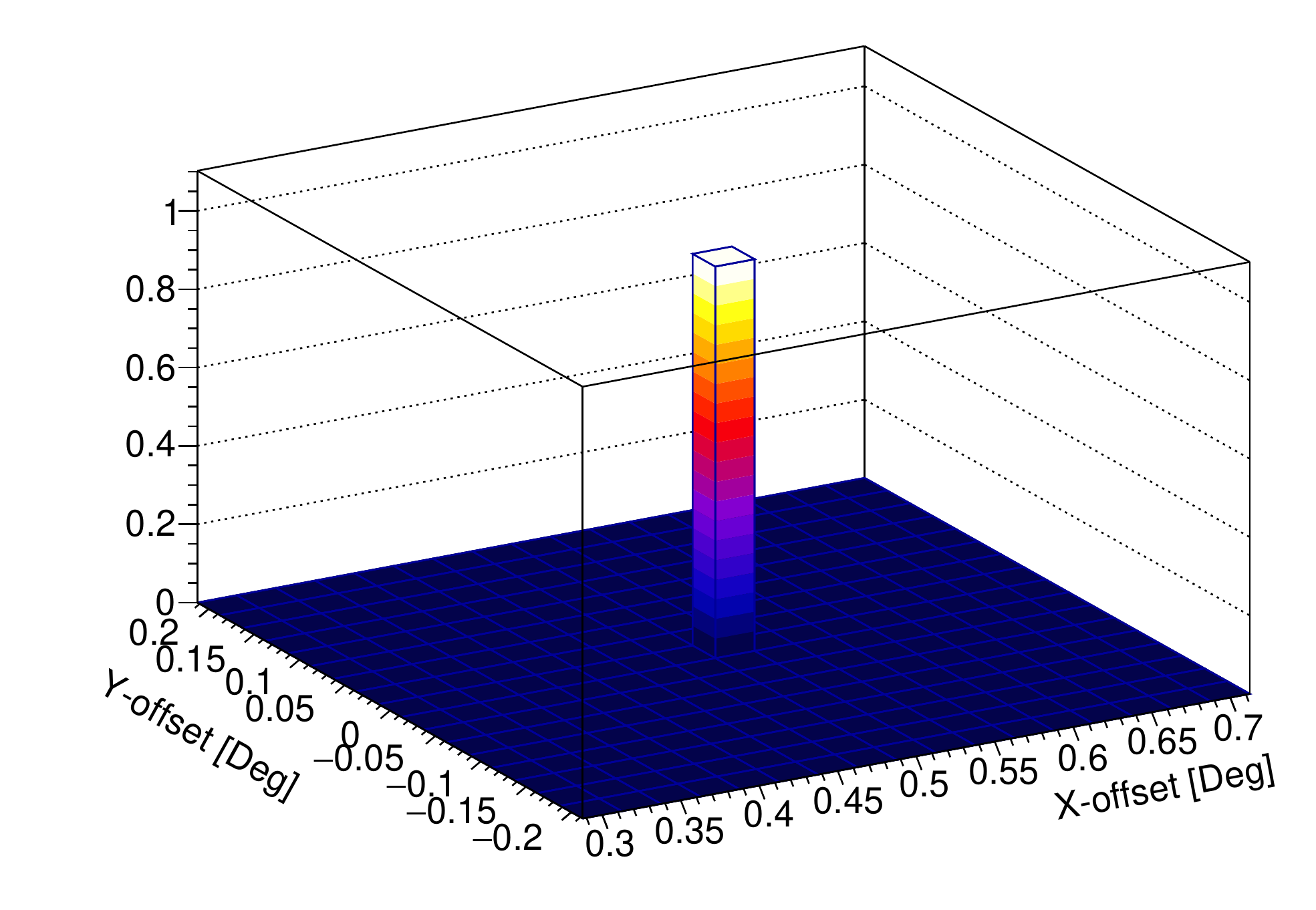}
      \caption{Input point source model}
      \label{fig:InitialPointSourceSpatialModel}
   \end{subfigure}
   \begin{subfigure}[b]{0.4\textwidth}
      \includegraphics[width=\textwidth]{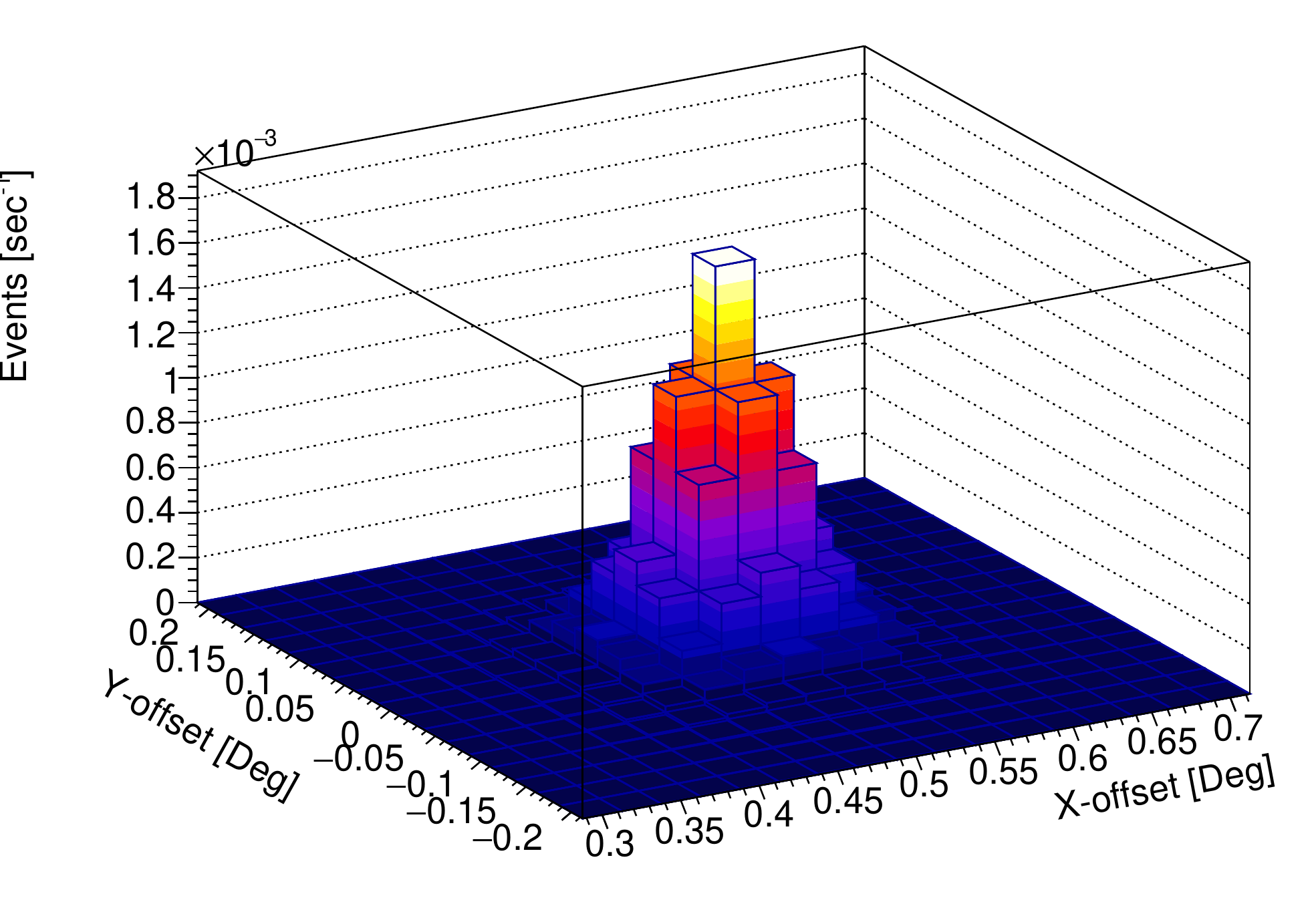}
      \caption{Computed point source model}
      \label{fig:FinalPointSourceSpatialModel}
   \end{subfigure}
   \caption{Example of a computed point source model using
     Eq. \protect\ref{eq:MLMSourceModel}. (a) Input source model assuming
     a delta function at the source position
     ($\sim$0.5$^{\circ}$ offset from the camera center). (b) 
     Resulting source model as computed by
     the MLM described here. For this
     test case, a Crab-like source spectrum was used.}
\label{fig:PointSourceModelConstruction}
\end{figure}

\begin{figure}
   \centering
   \begin{subfigure}[b]{0.4\textwidth}
      \includegraphics[width=\textwidth]{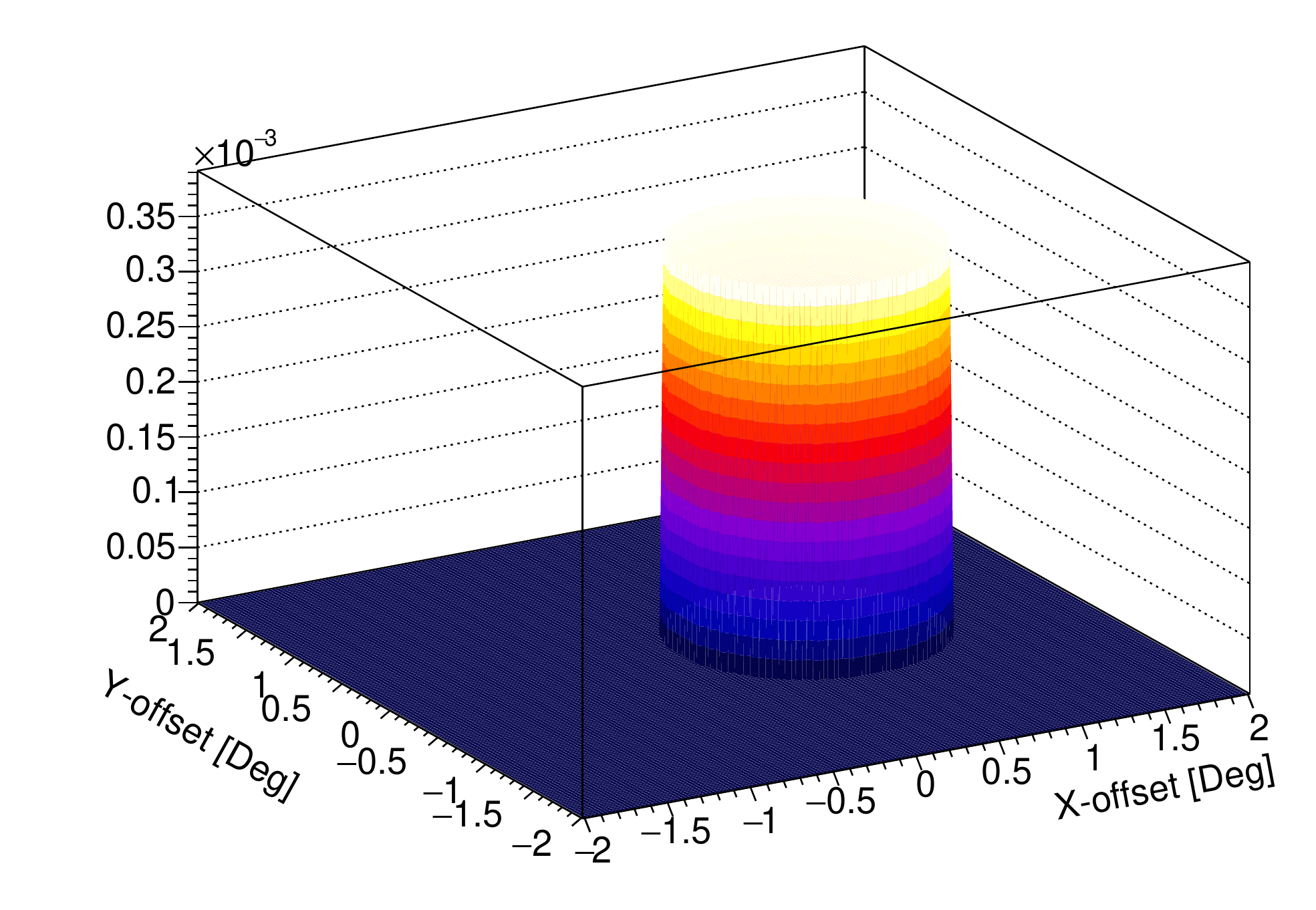}
      \caption{Input extended disc model}
      \label{fig:InitialExtendedSourceSpatialModel}
   \end{subfigure}
   \begin{subfigure}[b]{0.4\textwidth}
      \includegraphics[width=\textwidth]{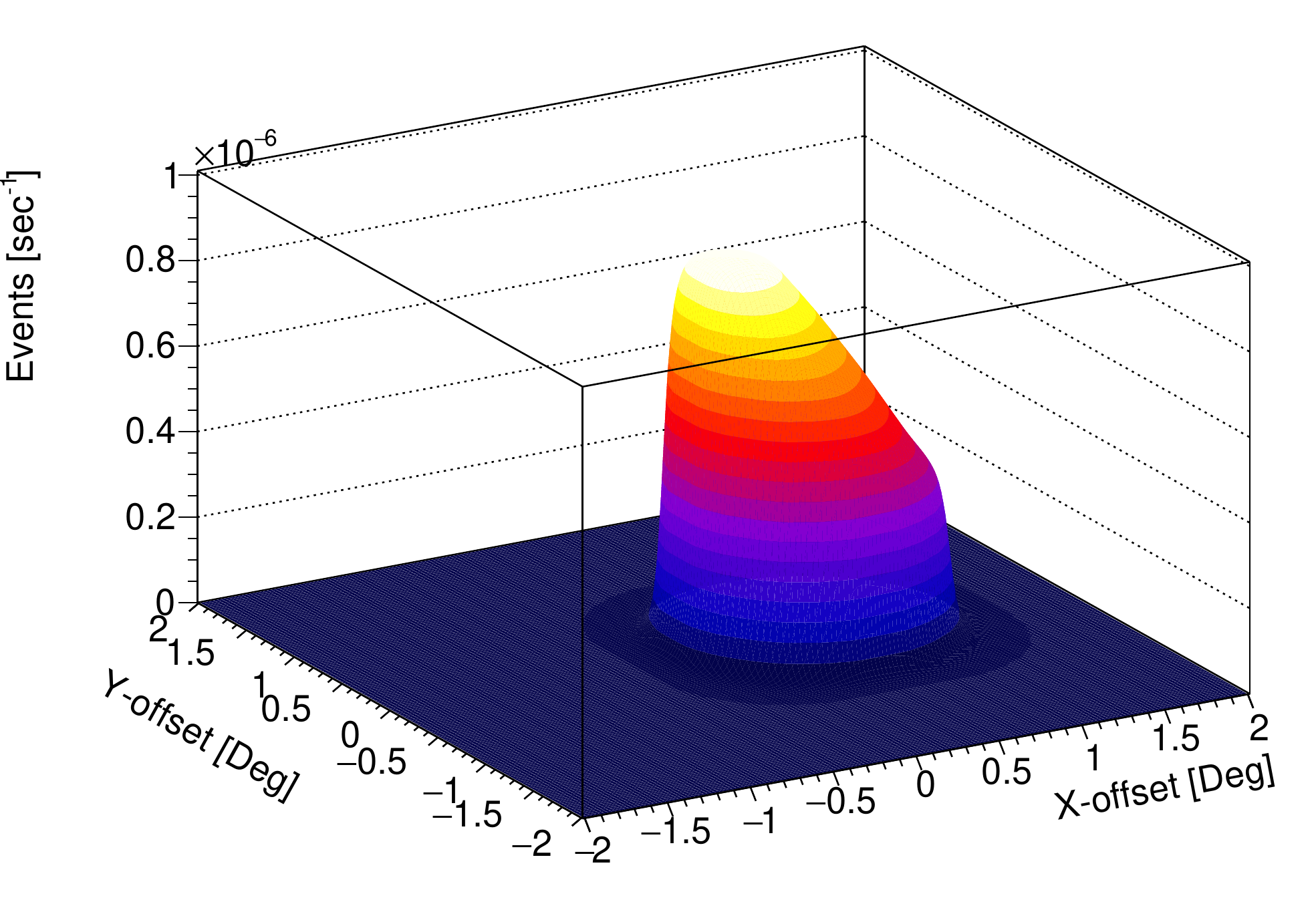}
      \caption{Computed extended disc model}
      \label{fig:FinalExtendedSourceSpatialModel}
   \end{subfigure}
   \caption{Example of a computed extended source model using
     Eq. \protect\ref{eq:MLMSourceModel}. (a) Input source model assuming
     a 1.5$^{\circ}$ diameter disc centered at 0.5$^{\circ}$ offset
     from the camera center. (b) Resulting source model as computed
     by the MLM. For this
     test case, a 10\% Crab-like source spectrum was used.}
\label{fig:ExtendedSourceModelConstruction}
\end{figure}

In order to produce a $\gamma$-ray source spatial model the
formulation from Mattox et al. (1996) Eq. 2 is used
\cite{1996ApJ...461..396M}.
\begin{equation}
\label{eq:MattoxEquation}
S_{src}(\mathbf{\vec{r}}|\mathbf{\vec{s}})=\frac{1}{N}\int_{E_{min}}^{E_{max}}\int_{0}^{\infty}[B(\mathbf{\vec{r}},
E')
\ast P(\mathbf{\vec{r}},E')] S(E'|\mathbf{\vec{s}}) R(\mathbf{\vec{r}},E',E) A(\mathbf{\vec{r}},E') dE'dE
\end{equation}
This gives the expected distribution of the source emission as it is
observed. \textit{B} represents the intrinsic source
morphology, which in the case of a point source is a delta
function centered at the source position. The various instrument
response functions (IRFs) accounted for are the point spread function
(PSF, \textit{P}), energy dispersion (\textit{R}), and effective
collection area (\textit{A}). Additional parameters are the intrinsic
source energy spectrum (\textit{S}), the spatial coordinates in the
field of view (\textit{$\mathbf{\vec{r}}$}), the true $\gamma$-ray energy
(\textit{E'}), and the observed $\gamma$-ray energy (\textit{E}). To
make the computation of
this model tractable, certain assumptions are employed. These include finely
binning \textit{B} and \textit{P} spatially
as well as finely binning in \textit{E'} \footnote{\textit{B} and
  \textit{P} are binned spatially using
  0.025$^{\circ}$$\times$0.025$^{\circ}$ bins. \textit{E'} is binned in
$log_{10}$(\textit{E'} [TeV]) at increments of 0.05.}. The spatial model then
becomes a summation rather than an integration
\begin{equation}
\label{eq:MLMSourceModel}
S_{src}(\mathbf{\vec{r}}_{i,j}|\mathbf{\vec{s}})=\sum_{k,m,n}B_{m,n}P_{m,n}(\mathbf{\vec{r}}_{i,j})A_{m,n}(E_{k}')\int_{E_{min}}^{E_{max}}R_{m,n}(E,E_{k}')dE\int_{E'_{k,lower}}^{E'_{k,upper}}S(E'|\mathbf{\vec{s}})dE'.
\end{equation}
This now gives the value of the spatial model within a given bin \textit{i,j} of
the spatial map. Potential ``smearing'' of events due to
the instrument resolution is accounted for by summing the contributions from
all other bins \textit{n,m} at position \textit{i,j} in the spatial
map. A pictoral example of the input source model and resulting MLM
computed source model for both a point source and 1.5$^{\circ}$
diameter extended disc source
are given in Figures \ref{fig:PointSourceModelConstruction} and
\ref{fig:ExtendedSourceModelConstruction} respectively.

This formulation solves an additional challenge. Following
Eq. \ref{eq:MattoxEquation} the entire source model and
all relavent contributions from each IRF need to be
recomputed at each step of the fit. This is computationally intensive. However, using
Eq. \ref{eq:MLMSourceModel} the component of the source model derived
from the IRFs (which are independent of the source spectral
parameters) can be precomputed for every bin of \textit{E'}
requiring only a reweighting of these bins for each iteration of the
fit.

\begin{figure}
  \begin{center}
   \includegraphics[width=0.5\textwidth]{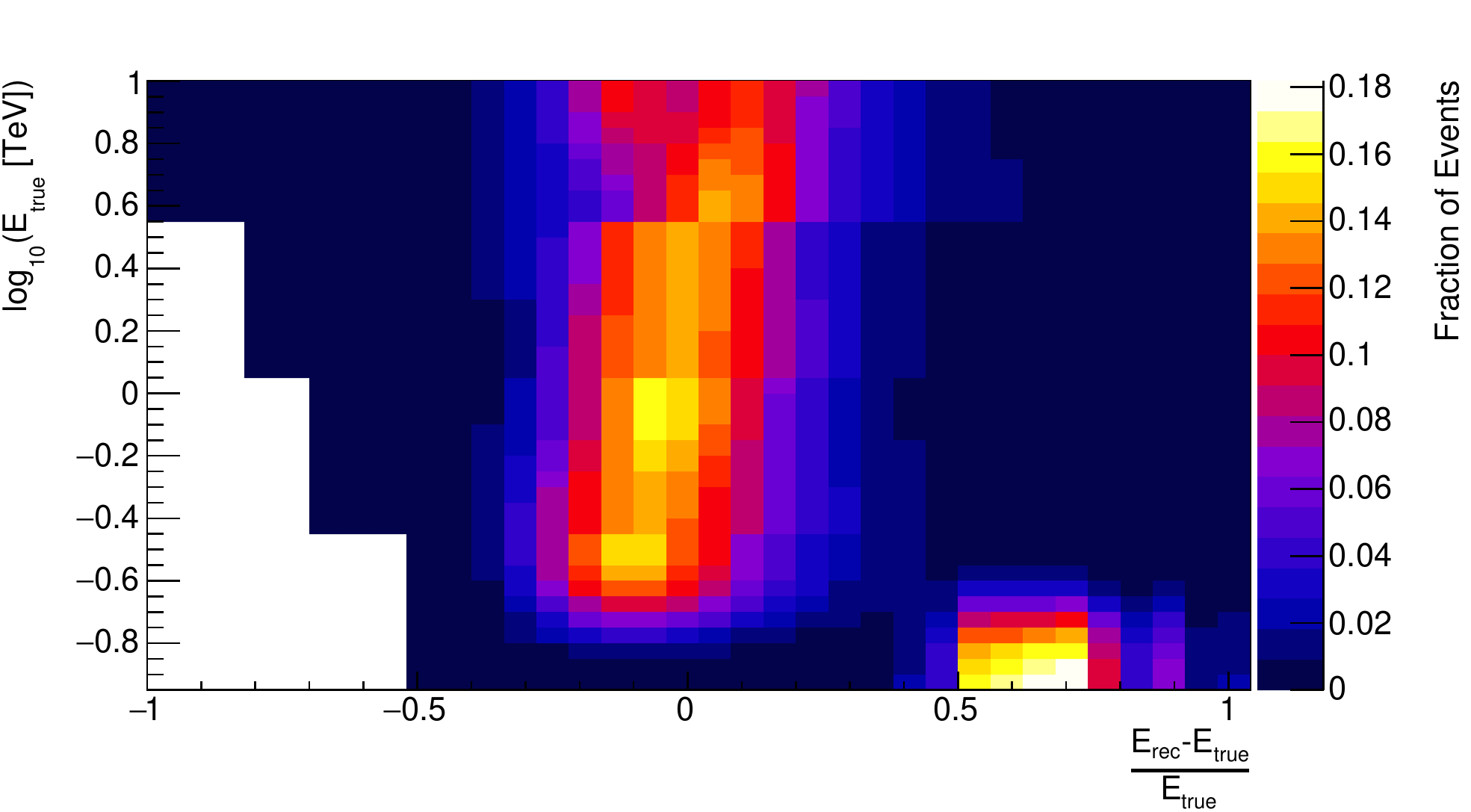}
  \end{center}
   \caption{Energy dispersion as a function of logarithmic true
     energy. The plot represents the fraction of $\gamma$-ray events
     reconstructed at a given energy (E$_{rec}$) as
     a function of (E$_{rec}$-E$_{true}$)/E$_{true}$. Values are based
     on fixed-energy $\gamma$-ray
     air-shower simulations at 70$^{\circ}$ elevation,
     180$^{\circ}$ azimuth, and offset by 0.25$^{\circ}$ from the
     camera center. Simulations are processed using non-standard cuts
     on MSW and including all events within the field of view. Values
     are interpolated at energies not represented by the
     fixed-energy simulations.}
   \label{fig:EnergyDispersion}
\end{figure}

The IRFs are parameterized for
a range of observing conditions including zenith angle, azimuth,
night sky noise level, observing offset from the source, atmospheric
conditions, telescope array configuration, and number of
telescopes contributing to the shower reconstruction. The PSF
is modeled based on fits of a King function to simulated $\gamma$-ray
air-shower
events generated at fixed energies of 0.1, 0.3, 1, 3, and 10 TeV
across the full range of observing conditions listed above.
\begin{equation}
\label{eq:King}
PSF(x,y)\propto\left(1-\frac{1}{\lambda}\right)\left(1+\frac{1}{2\lambda}\cdot\frac{x^{2}+y^{2}}{\sigma^{2}}\right)^{-\lambda}
\end{equation}
In particular $\lambda$ and $\sigma$ are computed for each combination
of possible observing conditions and interpolated on to obtain a
PSF model at any potential observing position and energy. These
simulations are also used to generate the energy dispersion of
the instrument. As the interpolated energy dispersion becomes less
well defined below $\sim$300 GeV (due to limited statistics and the
interpolation method used, see Figure \ref{fig:EnergyDispersion}) the MLM is currently restricted to analysis
above this energy.

\subsection{Mean Scaled Width}
\begin{figure}
\centering
        \begin{subfigure}[b]{0.49\textwidth}
                \includegraphics[width=\textwidth]{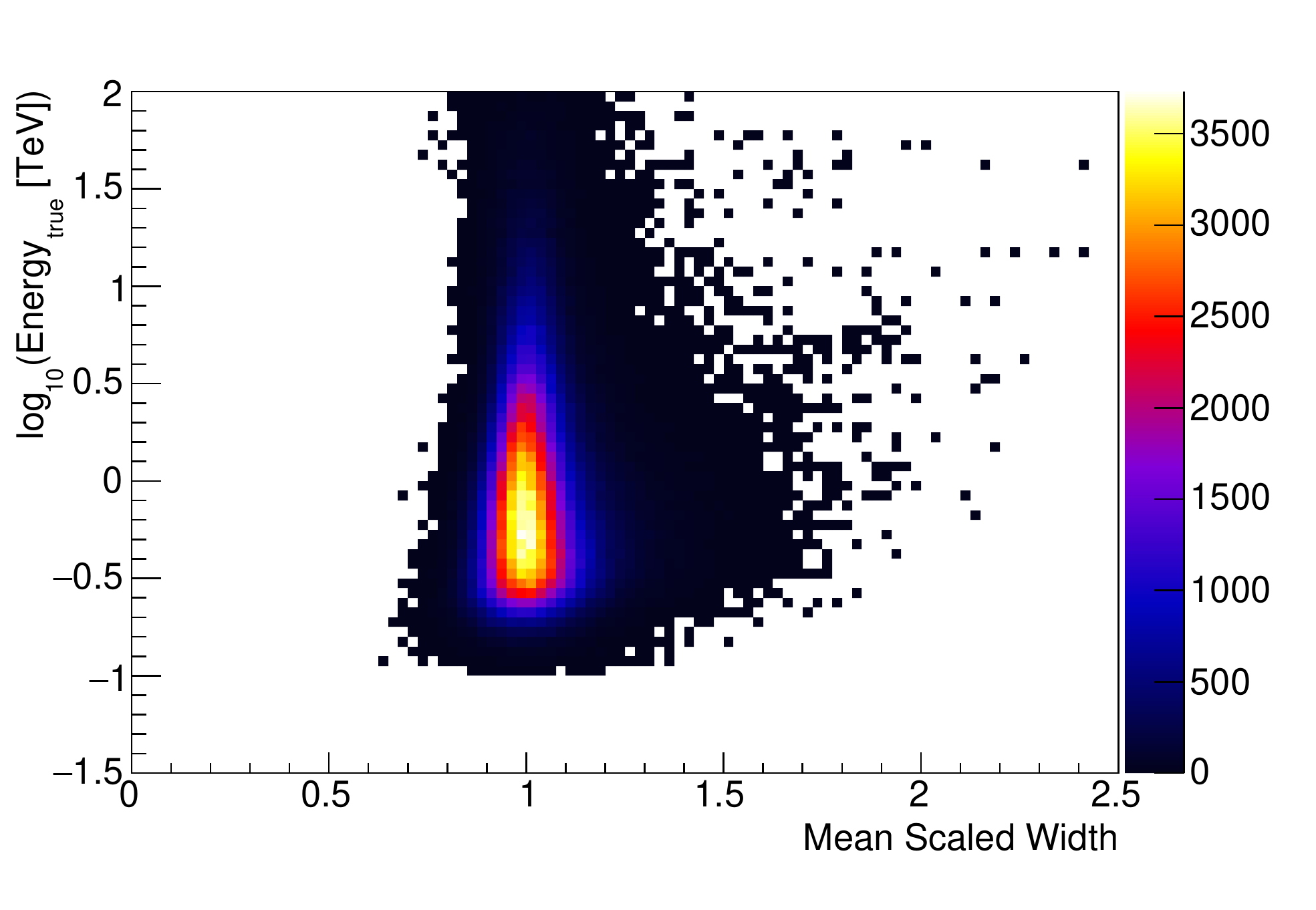}
                \caption{Gamma-ray MSW Distribution}
                \label{fig:SourceMSW}
        \end{subfigure}
        ~ 
        \begin{subfigure}[b]{0.49\textwidth}
                \includegraphics[width=\textwidth]{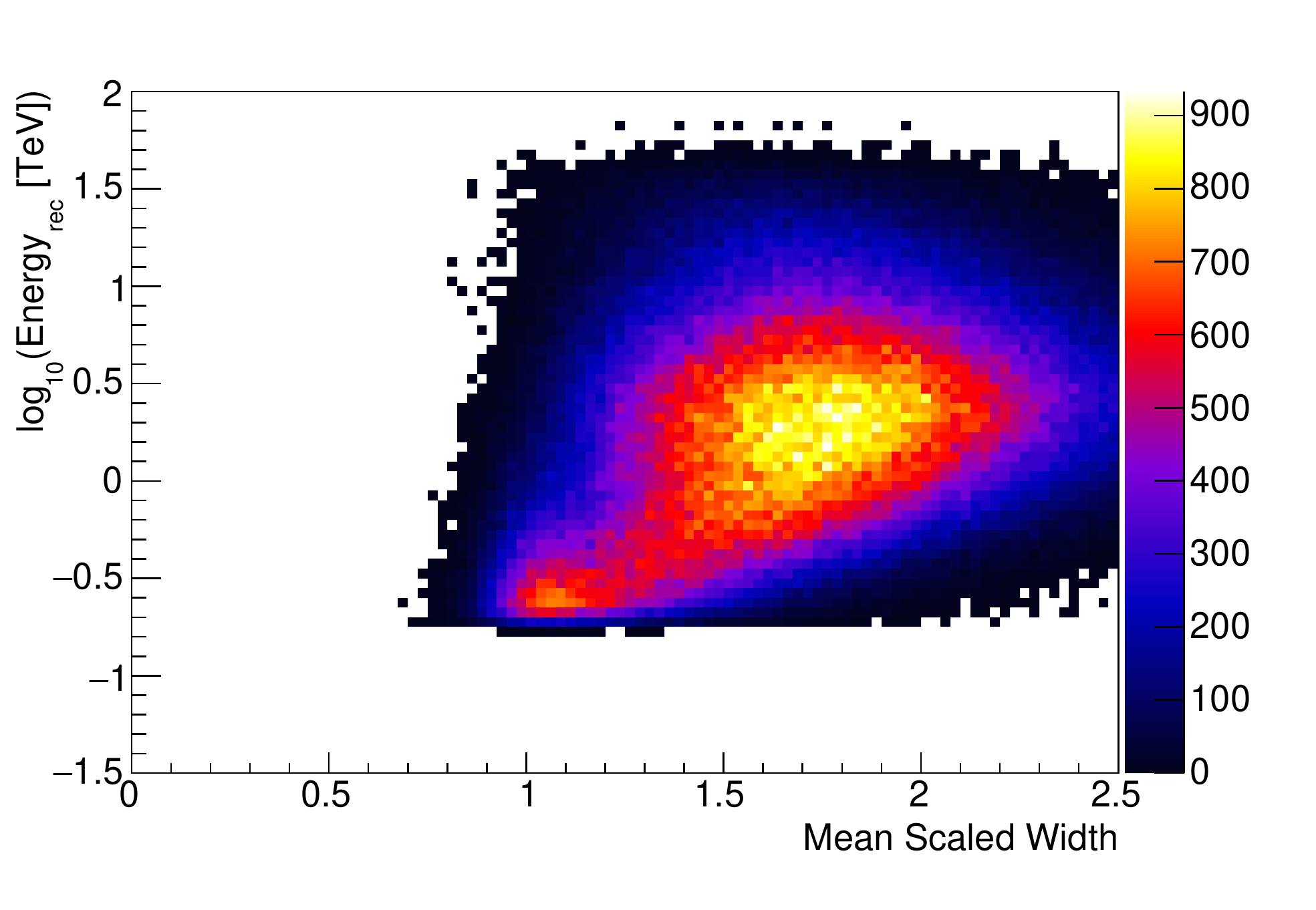}
                \caption{Background MSW Distribution}
                \label{fig:BackgroundMSW}
        \end{subfigure}
\caption{MSW distributions derived (a) for the source model from
  simulated $\gamma$-ray air-showers and (b) for the background model
  from observations. Distributions are shown as a function of
  logarithmic energy. The color scale represents the number of events
  used to generate the distribution. Models
  are extracted for an observation position of 70$^{\circ}$
  elevation, 180$^{\circ}$ azimuth, and 0.5$^{\circ}$ offset with four
telescopes participating in the shower reconstruction. The slight peak
near 1 in the background MSW distribution below
$\sim$300 GeV is currently under investigation.}
\label{fig:MSWDistributions}
\end{figure}

In order to improve sensitivity to extended sources a third
dimension based on a $\gamma$/hadron discriminating
parameter known as mean scaled width (MSW)
\cite{2008ICRC....3.1325D} is incorporated. This parameter is based on the average of
the Hillas width parameter \cite{1985ICRC....3..445H} from images of
all participating telescopes normalized by the expected width derived
from simulated $\gamma$-ray air-showers.
\begin{equation}
MSW=\left(\frac{1}{n}\right)\sum_{i=1}^{n}\frac{w_{i,j}}{<w_{i}>_{j}}
\end{equation}
\textit{n} is the number of participating telescopes and
\textit{j} represents the various shower parameters over which the
expected width value is determined, such as integrated image intensity
and impact distance. It follows that $\gamma$-ray events will have a
MSW distribution which peaks very close to 1. However, cosmic-rays, which
typically result in less compact images, will peak at higher values of
MSW (see Figure \ref{fig:MSWDistributions}). For the MLM analysis, MSW
is restricted to the range 0.8 - 1.3 to provide ample distinction
between the source and background models.

The source MSW model is derived from simulations of $\gamma$-ray
air-showers while the background models are derived from actual 
observations, as mentioned above\footnote{As there is potentially 
  a correlation between MSW and spatial position, 
  both sets of MSW distributions are derived for a range of different offsets.}.

\section{Validation Plans}\label{ValidationPlans}
The planned validation of the VERITAS MLM analysis will initially
involve a self consistency check in which the generated models are used to
produce a sample of toy Monte Carlo events. These generated events
will then be fit with the same models to ensure the fitted result is
consistent with inputs. The next step will be to fit the models to
dedicated \textit{OFF} observations to check the
background models are consistent with data. Once the background models
have been verified, a test on the Crab Nebula, a known bright point source,
will be conducted to check the source model generation is
accurate. Additionally, data has been taken on the Crab Nebula
following a raster scan pattern to produce an extended source
\cite{2015ICRC_CrabRasterScan}. Tests
on these datasets will provide a check of the extended source fitting
using a source with a known spectrum. Finally, the technique will be
applied to known extended
sources such as supernova remnant G106.3+2.7
\cite{2009ApJ...703L...6A} and MGRO J1908+06
\cite{2014ApJ...787..166A}. Each check is designed to ensure all
components of the MLM are understood and behave as expected as well as
to identify regions where the technique struggles.

\section{Discussion}
The expected number of $\gamma$-ray events predicted by the MLM for
the Crab-like source model presented in Figure
\ref{fig:PointSourceModelConstruction} was compared to a sample of
observations of the Crab nebula taken under similar observing
conditions and analyzed with a standard RBM analysis. The MLM
prediction was found to be consistent to within
$\sim$10\%. Additionally, the extended source (which uses a 10\%
Crab-like spectrum) actually predicts $\sim$8-9\% of the expected
Crab flux. This is due to the reduced instrument sensitivity at larger
offsets from the center of the field of view which is reflected in
the asymmetry of the computed model (see Figure
\ref{fig:FinalExtendedSourceSpatialModel}). 

The analysis is still under development and will undergo a set of
optimizations in the future. It will be tested whether
extending the range on MSW provides better characterization of the
background and thus improved separation between source and background
model components. Additionally, how the chosen coarse binning in
observed energy affects the result will be investigated. Improvements to
the VERITAS shower and energy reconstruction methods
will also lead to more accurate IRF parameterizations and thus improve
the overall MLM. 

Once validated, the technique can be
employed on several known extended sources currently undetected by
IACTs, such as the cocoon in Cygnus and MGRO J0632+17. This
technique also allows the possibility for doing
energy-dependent morphology fitting across multiple instruments using
a set of common spatial templates. In so doing the VERITAS MLM could
be an invaluable tool for multi-wavelength studies in the future.

\acknowledgments
This research is supported by grants from the U.S. Department of
Energy Office of Science, the U.S. National Science Foundation and the
Smithsonian Institution, and by NSERC in Canada. We acknowledge the
excellent work of the technical support staff at the Fred Lawrence
Whipple Observatory and at the collaborating institutions in the
construction and operation of the instrument. The researchers would
also like to thank our collaborators at DESY for their aid in
generating the mono-energetic simulations which served as inputs to
the PSF and energy dispersion IRFs.

The VERITAS Collaboration is grateful to Trevor Weekes for his seminal
contributions and leadership in the field of VHE gamma-ray
astrophysics, which made this study possible.

\bibliographystyle{JHEP}
\bibliography{Cardenzana_MLMProc_ICRC2015}{}

\end{document}